\documentclass[a4paper]{article}

\usepackage{INTERSPEECH_v2}

\usepackage{algorithm,algorithmic,cite,amsmath,amssymb,epsfig,enumerate,amsfonts,url,multirow,array,tabularx,booktabs,stfloats,flushend,diagbox,theorem}
\usepackage[T1]{fontenc}

\def\x{{\mathbf x}}
\def\w{{\mathbf w}}
\def\y{{\mathbf y}}

\def\z{{\mathbf z}}

\def\X{{\mathbf X}}
\def\W{{\mathbf W}}

\title{An Investigation of Universal Background Sparse Coding Based Speaker Verification on TIMIT}
\name{~Xiao-Lei~Zhang}
\address{
Center for Intelligent Acoustics and Immersive Communications, School of Marine Science and Technology, Northwestern Polytechnical University, China }
\email{xiaolei.zhang@nwpu.edu.cn, xiaolei.zhang9@gmail.com}

\begin{document}

\maketitle
\begin{abstract}
%\boldmath
In this paper, we propose a universal background model, named universal background sparse coding (UBSC), for speaker verification. The proposed method trains an ensemble of clusterings by data resampling, and produces sparse codes from the clusterings by one-nearest-neighbor optimization plus binarization. The main advantage of UBSC is that it does not suffer from local minima and does not make Gaussian assumptions on data distributions. We evaluated UBSC on a clean speech corpus---TIMIT. We used the cosine similarity and inner product similarity as the scoring methods of a trial. Experimental results show that UBSC is comparable to Gaussian mixture model.
\end{abstract}

\section{Introduction}

Speaker verification has long been a fundamental task in speech processing. In speaker verification, the recognizer verifies an identity claim made by a test speaker, and decides to accept or reject the claim. Based on the input speech material, speaker verification can be either \textit{text-dependent} or \textit{text-independent}. The former constrains the speaker to pronounce a prescribed text, while the latter does not constrain the speech contents. Here we study text-independent speaker verification.

The first and earliest speaker verification method, i.e. feature averaging, learns the utterance-level feature of an utterance by averaging the frame-level acoustic features \cite{markel1977long}. The method requires long speech utterances to reach stable speech statistics.

The second method estimates the density of speech frames by statistical models. Early approaches of this kind build a model, e.g. vector quantization \cite{soong1987report} or Gaussian mixture model (GMM) \cite{reynolds1995speaker,reynolds1995robust}, for each training speaker. These approaches are inefficient when the number of training speakers is large. To alleviate this problem, GMM-based universal background model (GMM-UBM), which builds a single GMM from the pool of all training speakers, was developed \cite{reynolds2000speaker}. GMM-UBM is a fundamental method of the later research. To deal with noise factors, such as utterance variations and channel variations, many approaches were proposed along with GMM-UBM, where i-vectors \cite{kenny2007joint,dehak2011front} are among the effective ones. They first extract high-dimensional supervectors from the first- and second-order statistics of GMM-UBM, and then reduce the noise factors by factor analysis.

The third method is based on deep neural networks (DNNs). It can be roughly categorized to two approaches. The first approach uses a DNN to extract bottleneck features that are then used as the input of GMM-UBM, e.g. \cite{sarkar2014combination}. The second approach takes a DNN trained for a different task, e.g. speech recognition, to generate class posteriors of speech frames \cite{lei2014novel}, which is an alternative to GMM-UBM. To demonstrate the advantages of the two approaches, their DNNs need to be trained with additional data \cite{richardson2015deep}.

After feature extraction by the aforementioned methods, speaker verification needs to score the similarity of two speakers in a trial.
The scoring methods include maximum a posteriori estimation \cite{reynolds2000speaker}, support vector machines \cite{campbell2006support}, cosine similarity measurement \cite{dehak2011front}, probabilistic linear discriminative analysis \cite{kenny2010bayesian}, etc.

Besides the aforementioned methods, sparse coding is another emerging topic \cite{imran2010sparse,kua2011speaker,li2011speaker,haris2015support,kua2013vector,boominathan2012speaker}. Many sparse coding methods focus on post processing \cite{imran2010sparse,kua2011speaker,li2011speaker,haris2015support,kua2013vector}, including modeling and classification of speakers with GMM-UBM supervectors and i-vectors. The orthogonal matching pursuit (OMP) method proposed by Boominathan and Murty \cite{boominathan2012speaker} is  an alternative to GMM-UBM, which directly builds a sparse model for each test utterance from original acoustic features, and produces a sparse code for each speech frame.

%To summarize, GMM-UBM \cite{reynolds2000speaker} is an important component of speaker verification. However, it assumes that data follows Gaussian distribution, which may not be always accurate. Its training method, i.e. expectation-maximization, also suffers from local minima.

In this paper, we propose a new UBM, named universal background sparse coding (UBSC), which directly builds a UBM from original acoustic features by data resampling and one-nearest-neighbor optimization. We compared UBSC with GMM-UBM on TIMIT. To analysis the difference between GMM-UBM and UBSC, we used neither denoising frontend nor additional data sets, and took the simple \textit{cosine similarity} measurement as the scoring method of a trial. Experimental results show that UBSC is comparable to GMM-UBM.

\section{Universal background sparse coding}

\subsection{Model training}

UBSC trains an ensemble of $k$-centers clusterings. The centers of a $k$-centers clustering is trained simply by random sampling. Specifically, suppose we have a number of training speakers, and each speaker contains several utterances. The training process is as follows:

 \begin{itemize}
 \item The first step extracts frame-level acoustic features, e.g. mel-frequency cepstral coefficients (MFCC), from the speech signals, and then pools all acoustic features together into a set, denoted as $\mathcal{X} = \{\x_i\}_{i}$, where $\x_i$ denotes the acoustic feature of the $i$-th frame.
 \item The second step builds $V$ random models $\{\W_v\}_{v=1}^V$, where the model $\W_v$ is a random sample of $k$ frames from $\mathcal{X}$ without replacement\footnote{The word ``without replacement'' means that the $k$ frames are different observations in $\mathcal{X}$.}, denoted as $\W_v=[\w_{v,1},\ldots,\w_{v,k}]$.
\end{itemize}

From the above, we can see that UBSC has two hyperparameters $k$ and $V$.

\subsection{Sparse representation learning}
In the feature learning stage, given an utterance $\X=[\x_1,\ldots,\x_N]$ of a speaker to be processed in either the training, enrollment, or test stage, a high-dimensional supervector $\z$ is extracted from $\X$ by Algorithm \ref{alg:1}.

\begin{figure}[t]
 \centering
         \includegraphics[width=3in]{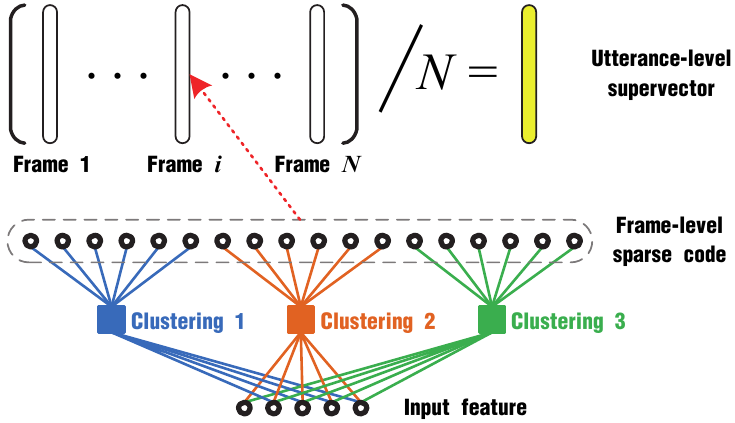}
         \caption{Principle of UBSC. The three base clusterings are drawn in different colors.}
 \label{fig:principle}
 \end{figure}

Algorithm \ref{alg:1} first calculates frame-level sparse features $\{\bar{\mathbf{s}}_i\}_{i=1}^N$ and then averages the frame-level features for an utterance-level supervector $\z$, where $\bar{\mathbf{s}}_i$ is the concatenation of a group of one-hot codes $\{\mathbf{s}_{i,v}\}_{v=1}^V$, each of which is produced from a random model $\W_v$. The learning process is illustrated in Fig. \ref{fig:principle}.

 \begin{algorithm}[t]
    \caption{}
    \begin{algorithmic}[1]\label{alg:1}
\REQUIRE  Input utterance $\X=[\x_1,\ldots,\x_N]$ and UBSC model $\{\W_v\}_{v=1}^V$ where $\W_v=[\w_{v,1},\ldots,\w_{v,k}]$
\ENSURE High-dimensional supervector $\z$
\FOR{$i=1,\ldots,N$}
    \FOR{$v=1,\ldots,V$}
        \FOR{$j=1,\ldots,k$}
            \STATE  $d_{i,v,j} \leftarrow \|\x_i-\w_{v,j}\|_2$.
        \ENDFOR
        \STATE Learn a sparse code $\mathbf{s}_{i,v} = [s_{i,v,1},\ldots,s_{i,v,k}]^T$ by
        \begin{equation}\label{eq:alg}
          s_{i,v,j} \leftarrow \left\{\begin{array}{ll}
            1,\mbox{ if }d_{i,v,j}=\min_{\forall l\in\{1,\ldots,k\}} d_{i,v,l}\quad\\
            0,\mbox{ otherwise}
          \end{array} \right.
        \end{equation}
    \ENDFOR
    \STATE $\bar{\mathbf{s}}_i\leftarrow [\mathbf{s}^T_{i,1},\ldots,\mathbf{s}^T_{i,V}]^T$
\ENDFOR
\RETURN $\z \leftarrow \frac{1}{N}\sum_{i=1}^N \bar{\mathbf{s}}_i$
\end{algorithmic}
\end{algorithm}

\section{Motivation and related work}\label{sec:system}
UBSC was motivated from multilayer bootstrap networks \cite{zhang2014multilayer}. They share the same theoretical base (see \cite{zhang2014multilayer} for the theoretical base of UBSC). However, UBSC may not be replaced by multilayer bootstrap networks. Because the scoring process of speaker verification is a supervised classification problem, learning a multilayer network in an unsupervised manner loses much information needed for the supervised problem. Empirically, we observed performance drop by using multilayer bootstrap networks.

UBSC is a comparable model to GMM-UBM. Comparing to GMM-UBM, UBSC does not make model assumptions on data distributions, since each base model of UBSC is a random sample of data. UBSC does not suffer from local minima, since it is optimized by one-nearest-neighbor. UBSC may be implemented easily. Its training process is also fast. A drawback of UBSC is that its network is usually large, so that its test complexity may be high. But it supports parallel computing naturally, which may alleviate this drawback.

UBSC is different from the OMP sparse coding method in \cite{boominathan2012speaker}. First, OMP learns a sparse model for each test utterance, while UBSC builds a single UBM from all training utterances. Second, OMP is formulated as an NP-hard combinatorial optimization problem, while UBSC is a simple algorithm that contains only data resampling and one-nearest-neighbor optimization. Third, OMP produces the test score of a trial directly, while UBSC produces a high-dimensional supervector for each utterance, leaving the scoring method an open problem. Besides, UBSC has a clear geometric explanation. It also supports parallel computing naturally.

% 介绍与GMM在概率描述上的差别

\section{Empirical results}
In this section, we compared UBSC with GMM-UBM on a clean speech corpus---TIMIT, and adopted a similar experimental setting with that in \cite{reynolds1995speaker}. All experiments were conducted with MATLAB 2015b on a Linux server running with 2 Inter(R) Xeon(R) E5-2650 CPUs, 4 Nvidia Tesla K80 GPUs (including 8 GPU cores), and 256 GB memory. Here we report the main results, leaving the details in the Supplementary Material in http://www.xiaolei-zhang.net/publications.htm.

\subsection{Experimental settings}
TIMIT contains 630 speakers, including 438 males and 192 females. Each speaker has 10 clean utterances. Each utterance is roughly 3 seconds long. The sampling rate of TIMIT is 16 kHz. To guarantee the reproducibility of the experiments, we did not adopt voice activity detection. We set frame length to 25 ms and frame shift to 10 ms. Then, we applied Hamming window filter to each frame, and extracted 19 dimensional MFCC with 1 dimensional log power energy by the VOICEBOX toolbox.\footnote{http://www.ee.ic.ac.uk/hp/staff/dmb/voicebox/voicebox.html} We further filtered the 20 dimensional features by a Hamming window in the mel-domain.

We adopted the MSR Identity Toolbox as the implementation of the GMM-UBM baseline.\footnote{https://www.microsoft.com/en-us/research/publication/msr-identity-toolbox-v1-0-a-matlab-toolbox-for-speaker-recognition-research-2/} In the training stage of GMM-UBM, we initialized the mean (and variance) of each Gaussian component by the mean (and variance) of the MFCC features of a randomly selected utterance, and set all Gaussian components to an equal prior probability. In the test stage, given an utterance, we extracted first- and second-order statistics from each frame \cite{reynolds2000speaker} which are further concatenated to a frame-level feature. Then, we averaged the frame-level features for an utterance-level supervector.

The parameter settings of the comparison methods are as follows.
For GMM-UGM, we searched the number of Gaussian mixtures from $2^{[1:1:11]}$ and the number of EM iterations from $[1,5,10,20,30]$ in grid, where the symbol $[a:b:c]$ represents a serial integers from $a$ to $c$ with an increment of $b$.
For UBSC, we searched parameter $k$ from $2^{[1:1:17]}$ and parameter $V$ from $[1,3,10,30]$ in grid. For each parameter pair, we recorded the average \textit{equal error rate} (EER) and standard deviation of 10 independent runs.

We used the two-tailed $t$-test to evaluate the statistical significance of the difference between a pair of results, where the null-hypothesis is defined as that the difference is insignificant. The significance level $\alpha$ is set to 0.05. If the $p$-value is smaller than $\alpha$, then the null-hypothesis is rejected, in other words, the difference is regarded as statistically significant.

\subsection{Results with cosine similarity scoring method}\label{subsec:result}
We used the \textit{cosine similarity} measurement as the scoring method of two supervectors \cite{dehak2011front}, which is defined by $\frac{\x^T\y}{{\|\x\|_2}{\|\y\|_2}}$ where $\x$ and $\y$ are two vectors.
 
 \subsubsection{Main results}\label{subsec:main}

For each speaker in TIMIT, we selected the first 8 utterances as training speech with each of the remaining 2 utterances as an individual test. We took each speaker as a claimant with the remaining speakers acting as imposters, and rotated through the tests of all speakers. We investigated the comparison methods on males, females, and both genders of speakers respectively. The number of claimant and imposter trials are summarized in Table \ref{tab:data}.

 \begin{table} [t]
\caption{\label{tab:data} {Number of claimant and imposter trials.}}
\centerline{
\scalebox{0.9}{
\begin{tabular}{lccc}
\hline
 & \#speakers & \#true trials  & \#imposter trials  \\
 \hline
Male& 438 & 876 & 765,624\\
Female& 192 & 384 & 146,688 \\
Male$+$Female& 630& 1,260 & 1,585,080 \\
\hline
\end{tabular}}
}
\end{table}

We list the best EERs of the comparison methods in Table \ref{tab:result0} with examples of the detection error tradeoff (DET) curves shown in Fig. \ref{fig:DET} and Figs. S1 and S2 in the Supplementary Material. From the comparison, we observe that UBSC shows statistically significant improvement over GMM-UBM in the Female experiment with a $p$-value of $0$, and slightly better performance than GMM-UBM in the Male and the Male$+$Female experiments with $p$-values of $0.2849$ and $0.0532$ respectively.

 \begin{table} [t]
\caption{\label{tab:result0} {EERs (in percent) produced by GMM-UBM and UBSC, when the cosine similarity scoring method is used. The values in brackets are standard deviations.}}
\centerline{
\scalebox{0.9}{
\begin{tabular}{lccc}
\hline
 & Male & Female  & Male$+$Female \\
 \hline
GMM-UBM& \textbf{3.92} (0.26) & 5.46 (0.55) & \textbf{3.35} (0.24)\\
UBSC& \textbf{3.79} (0.28) & \textbf{3.99} (0.23) & \textbf{3.16} (0.16) \\
\hline
\end{tabular}}
}
\end{table}

\begin{figure}[t]
 \centering
         \includegraphics[width=3in]{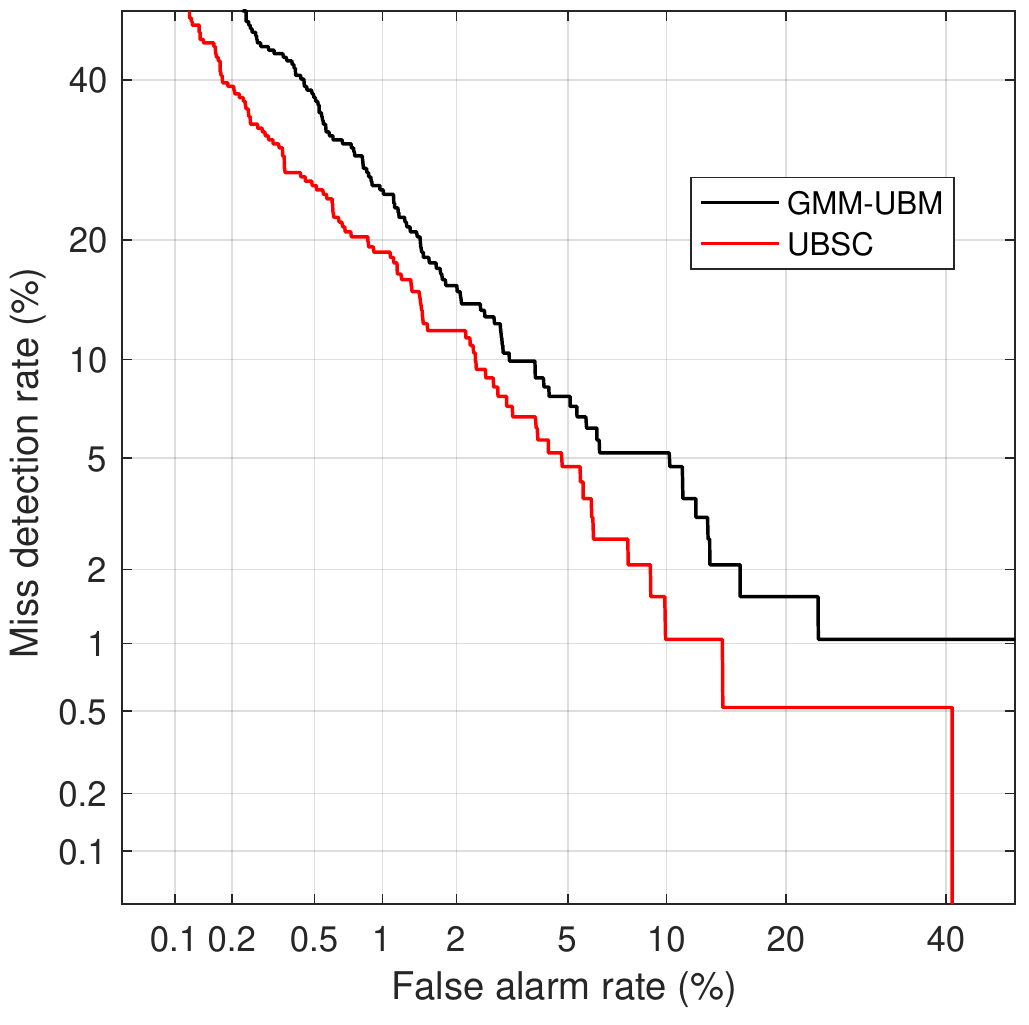}
         \caption{DET curves produced by GMM and UBSC in the Female experiment, with the cosine similarity scoring method.}
 \label{fig:DET}
 \end{figure}

\subsubsection{Effect of parameters $k$ and $V$}\label{subsec:parameter}
%\begin{figure}[t]
% \centering
%         \includegraphics[width=3in]{fig_parameter_v_k.pdf}
%         \caption{EER curves produced by the UBSC with the cosine similarity measurement with respect to parameters $V$ and $k$  in the Male$+$Female experiment.}
% \label{fig:MRCG_ROC}
% \end{figure}

We report the EER with respect to parameters $k$ and $V$ in Fig. \ref{fig:MRCG_ROC} and Figs. S3 and S4 in Supplementary Material. From the figures, we observe the following two phenomena. (i) If $k$ is fixed, then enlarging $V$ reduces EER, and moreover, setting $V$ to 30 balances the performance and computational complexity. (ii) Given $V$ fixed, we can find an optimal $k$.

  \begin{figure}[!htb]
 \centering
         \includegraphics[width=3in]{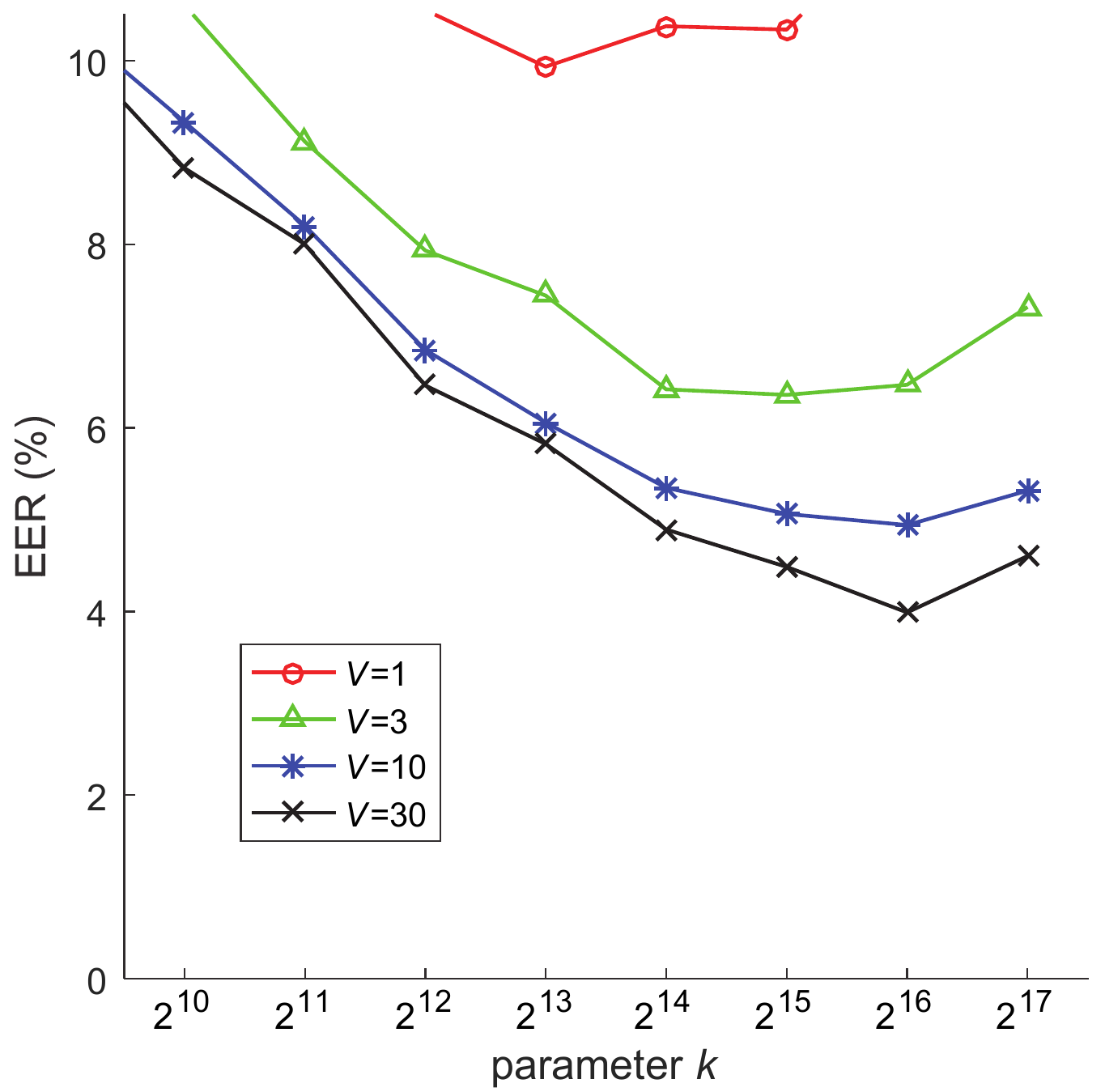}
         \caption{EER curves produced by the UBSC with the cosine similarity scoring method with respect to parameters $V$ and $k$ in the Female experiment.}
 \label{fig:MRCG_ROC}
 \end{figure}

\subsubsection{Effect of number of training speakers}\label{subsec:training}
\begin{figure*}[t]
 \centering
         \includegraphics[width=6.5in]{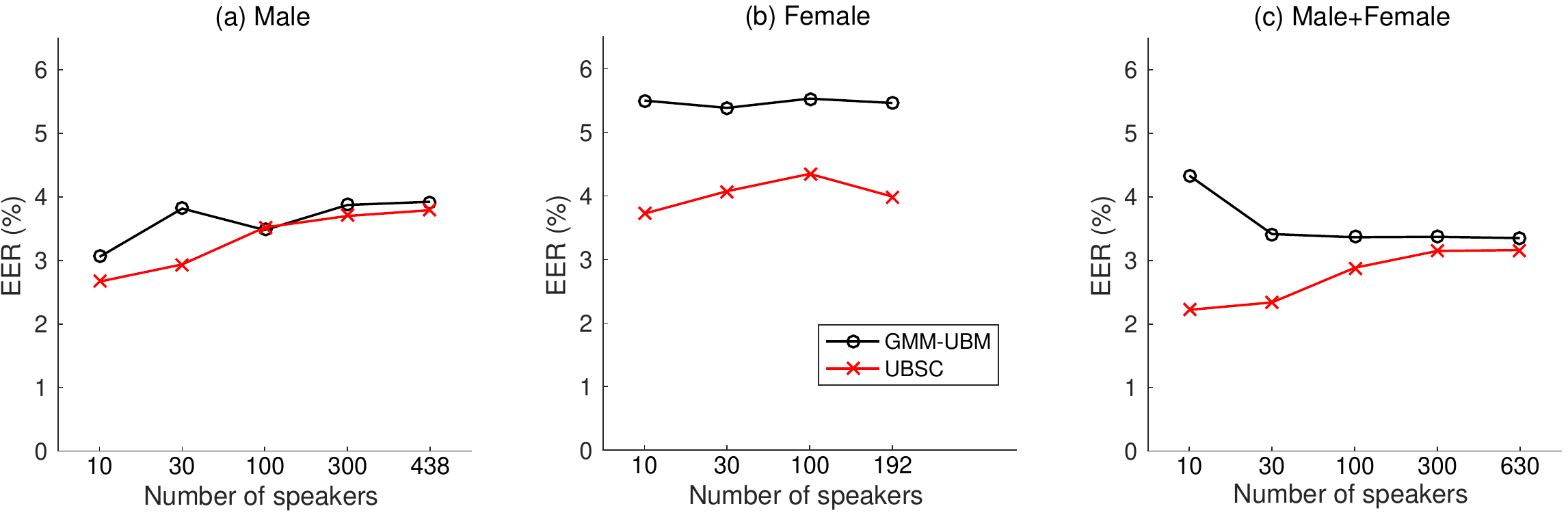}
         \caption{EER curve comparison of UBSC and GMM-UBM with respect to different number of training speakers, when the cosine similarity scoring method is used.}
 \label{fig:subset}
 \end{figure*}

To study how the number of training speakers affect the performance, we randomly select 10, 30, 100, and 300 speakers from males, females, and both genders respectively. We ran the experiment 10 times and report the average performance in Fig. \ref{fig:subset}. From the figure, we find that the empirical conclusions in Section \ref{subsec:main} are not affected by the number of training speakers.

Moreover, the advantage of UBSC with a small number of training speakers are more apparent than that with a relatively large number of training speakers. Possible explanations include that, when only a small number of training speakers are used, (i) the local minima of the expectation-maximization algorithm of GMM affects its performance heavily, and (ii) the data distribution may not be Gaussian.

 \begin{figure*}[t]
 \centering
         \includegraphics[width=6.5in]{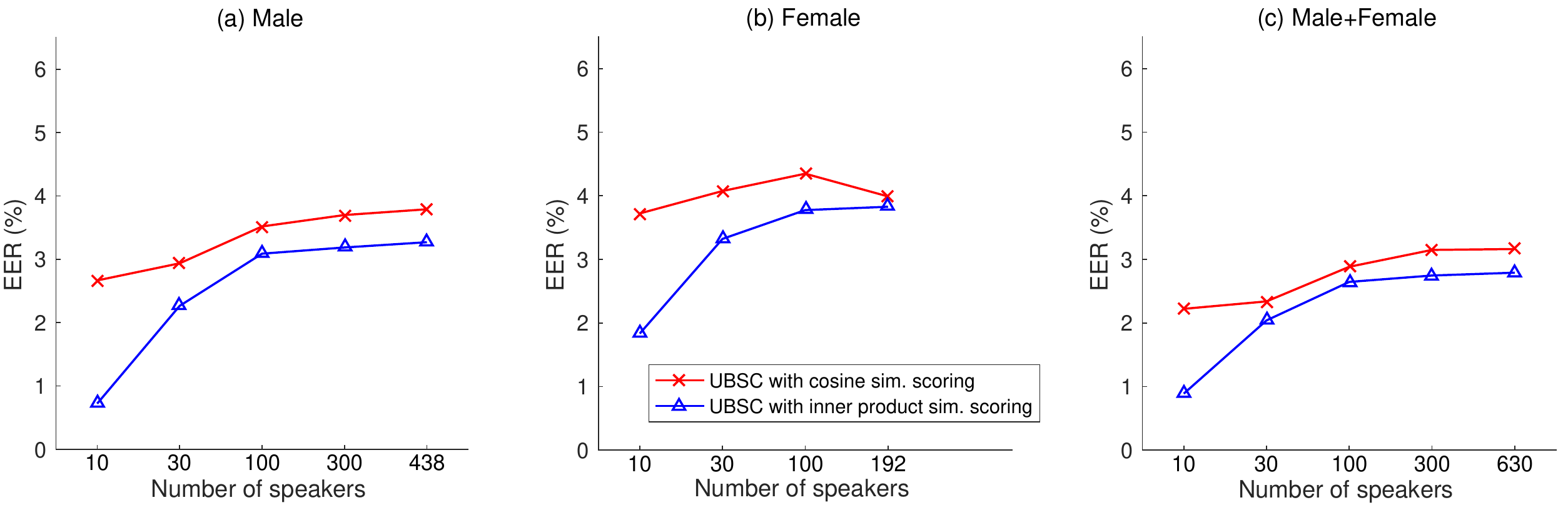}
         \caption{EER curve comparison between the UBSC with different scoring methods with respect to different number of training speakers.}
 \label{fig:subset_UBSCmetric}
 \end{figure*}

\subsection{Effect of scoring methods}
To investigate how scoring methods affect the performance of UBSC, we compared the {cosine similarity} scoring method with the \textit{inner product similarity} scoring method, where the inner product similarity is defined by ${\x^T\y}$.

 \begin{table} [t]
\caption{\label{tab:result1} {EERs (in percent) produced by UBSC. The values in brackets are standard deviations. The word ``sim.'' is short for similarity.}}
\centerline{
\scalebox{0.8}{
\begin{tabular}{lccc}
\hline
 & Male & Female  & Male$+$Female \\
 \hline
UBSC with cosine sim.& 3.79 (0.28) & \textbf{3.99} (0.23) & 3.16 (0.16) \\
UBSC with inner product sim.& \textbf{3.27} (0.21) & \textbf{3.83} (0.26) & \textbf{2.79} (0.18) \\
\hline
\end{tabular}}
}
\end{table}

We used the same experimental setting as that in Section \ref{subsec:main}.
We list the best EERs of the two scoring methods in Table \ref{tab:result1}. From the comparison, we observe that the inner-product similarity scoring method shows statistically significant improvement over the cosine similarity scoring method in the Male and the Male$+$Female experiments with $p$-values of $0.0001$ and $0.0001$ respectively, and is slightly better performance than the latter in the Female experiment with a $p$-value of $0.1539$.

To study how the number of training speakers affect the performance, we followed the same experimental setting as that in Section \ref{subsec:training}. We report the average performance in Fig. \ref{fig:subset_UBSCmetric}. From the figure, we find that the experimental conclusions on the two scoring methods of UBSC are not affected by the number of training speakers.

\section{Conclusions and future work}
In this paper, we have introduced a universal background model, called UBSC, for speaker verification. UBSC is trained simply by data resampling where each random sample of data forms the centers of a base clustering. In the test stage, given an utterance, UBSC first learns a frame-level sparse code by concatenating the one-hot output codes produced from the base clusterings, and then averages the frame-level sparse codes of all frames for an utterance-level supervector. UBSC does not make model assumptions and does not suffer from local minima. It is easily implemented and used. It supports parallel computing naturally.

We compared UBSC with GMM-UBM on the clean corpus---TIMIT. We used the cosine similarity and inner product similarity as the scoring methods. Experimental results show that, when the scoring method is the cosine similarity measurement, UBSC performs better than GMM on females, and is comparable to GMM on males and both genders of speakers. Moreover, the UBSC with the inner product similarity performs better than that with the cosine similarity. The conclusion is consistent with different number of speakers.

In the future, we will develop a denoising frontend and a backend classifier for UBSC. We will compress the scale of the UBSC model for speeding up the extraction of the supervectors. We will also investigate the density estimation ability of UBSC on more complicated data distributions.

\bibliography{zxlrefs,mywork}
\bibliographystyle{IEEEtran}

\end{document}